\documentclass[a4paper,fleqn]{cas-dc} %

\usepackage[authoryear,round]{natbib}

\usepackage{algorithm}
\usepackage{algpseudocode}

\usepackage{hyperref}
\hypersetup{
    colorlinks=true,         
    linkcolor=black,         
    citecolor=[rgb]{0.12, 0.47, 0.71}, 
    urlcolor=[rgb]{0.12, 0.47, 0.71},  
    anchorcolor=black
}

\def\tsc#1{\csdef{#1}{\textsc{\lowercase{#1}}\xspace}}
\tsc{WGM}
\tsc{QE}
\tsc{EP}
\tsc{PMS}
\tsc{BEC}
\tsc{DE}

\begin{document}
\let\WriteBookmarks\relax
\def\floatpagepagefraction{1}
\def\textpagefraction{.001}

\shorttitle{MOMAT: Mixture of Multiple Atlases for Low-Power Jailbreak Defense of Quantized LLMs }
\shortauthors{B. Li et~al.}

\title [mode = title]{
MOMAT: Mixture of Multiple Atlases for Low-Power Jailbreak Defense of Quantized LLMs  
}

\author[1]{Boyang Li}
\fnmark[1]
\cormark[1]
\ead{boli@kean.edu}
\author[2]{Bingyu Shen}
\fnmark[1]
\ead{bingyu.shen@hotmail.com}
\author[1]{Weihao Hong}
\fnmark[1]
\ead{hongw@kean.edu}
\author[1]{Zhiyuan Jiang}
\fnmark[1]
\ead{jianzhiy@kean.edu}
\author[1]{Xinlei Guan}
\fnmark[1]
\ead{guanxi@kean.edu}
\author[1]{Yan Ma}
\ead{yan.ma@kean.edu}
\author[4]{Miles Q. Li}
\ead{Qi.li7@mail.mcgill.ca}
\author[5]{Yi Sheng}
\ead{sheng1@usf.edu}
\author[3]{Ruiyang Qin}
\ead{ruiyang.qin@villanova.edu}

\cortext[cor1]{Corresponding author}
\fntext[fn1]{These authors contributed equally to this work.}

\affiliation[1]{organization={Department of Computer Science, Kean University},
                addressline={1000 Morris Ave},
                city={Union},
                postcode={07083},
                state={NJ},
                country={USA}}

\affiliation[2]{organization={Department of Computer Science and Engineering, University of Notre Dame},
                addressline={257 Fitzpatrick Hall of Engineering},
                city={Notre Dame},
                postcode={46556},
                state={IN},
                country={USA}}

\affiliation[3]{organization={Department of Computing Sciences, Villanova University},
                addressline={800 E Lancaster Ave},
                city={Villanova},
                postcode={19085},
                state={PA},
                country={USA}}

\affiliation[4]{organization={McGill University},
                addressline={845 Sherbrooke Street},
                city={Montréal},
                postcode={H3A 0G4},
                state={Québec},
                country={Canada}}

\affiliation[5]{organization={Department of Computer Science and Engineering, University of South Florida},
                addressline={4202 E Fowler Ave},
                city={Tampa},
                postcode={33620},
                state={FL},
                country={USA}}

\nonumnote{In this work, we show that restructuring safety
retrieval into domain-localized atlases and accelerating
similarity search with compute-in-memory hardware can
recover alignment robustness lost to quantization while
meeting the latency and energy constraints of edge
deployment. All correspondence regarding dataset
distribution or code reproduction should be directed to
the corresponding author.
  }

\begin{abstract}
Quantized large language models are increasingly deployed on edge devices for their low latency and energy efficiency. 
However, model quantization weakens alignment safeguards, leaving qLLMs (quantized large language models) highly vulnerable to jailbreak attacks. 
To address this challenge, we present MOMAT (Mixture of Multiple Atlases), a hardware-enhanced safety framework that combines structured knowledge retrieval with low-power defense acceleration.
Each atlas represents a semantic cluster of harmful or benign sample sets and policy templates, enabling domain-localized Retrieval-Augmented Generation guarding that mitigates the curse of dimensionality and the resulting semantic sparsity problem in large, heterogeneous safety databases.
MOMAT retrieves top-$k$ similarity features from all atlases for each prompt and evaluates them using a lightweight MoE (Mixture of Experts) detector, while a CiM (Compute-in-Memory)-accelerated similarity engine performs fast, low-power atlas-local retrieval.
MOMAT’s CiM-based retrieval accelerates a 100-query batch from 15,052.44 ms to 3,207.21 ns (a $4.69\times 10^6 \times$ speedup) and reduces energy from $8.1\times10^7 \mu$J to $3.32 \mu$J, yielding a ~$2.5\times 10^5 \times$ energy reduction over DRAM-based (Raspberry Pi) baselines.
Red-team evaluations across standard benchmarks show that MOMAT matches the defense performance of state-of-the-art methods while avoiding benign overkill and providing substantial efficiency gains, demonstrating that CiM-based modular defenses can make edge-deployed qLLMs both safer and more energy-efficient.
We will release the full 223.2k-sample dataset to foster future research.
\end{abstract}

\begin{keywords}
\sep Jailbreak Defense\sep \ Quantized Large Language Models \sep \ Retrieval-Augmented Generation \sep \ Compute-in-Memory \sep \ Mixture of Experts \sep \ Edge Deployment \sep \ AI Safety
\end{keywords}

\maketitle


\section{Introduction}
\label{sec:introduction}

\begin{figure*}[ht]
\begin{center}
 \includegraphics[width=0.8\linewidth]{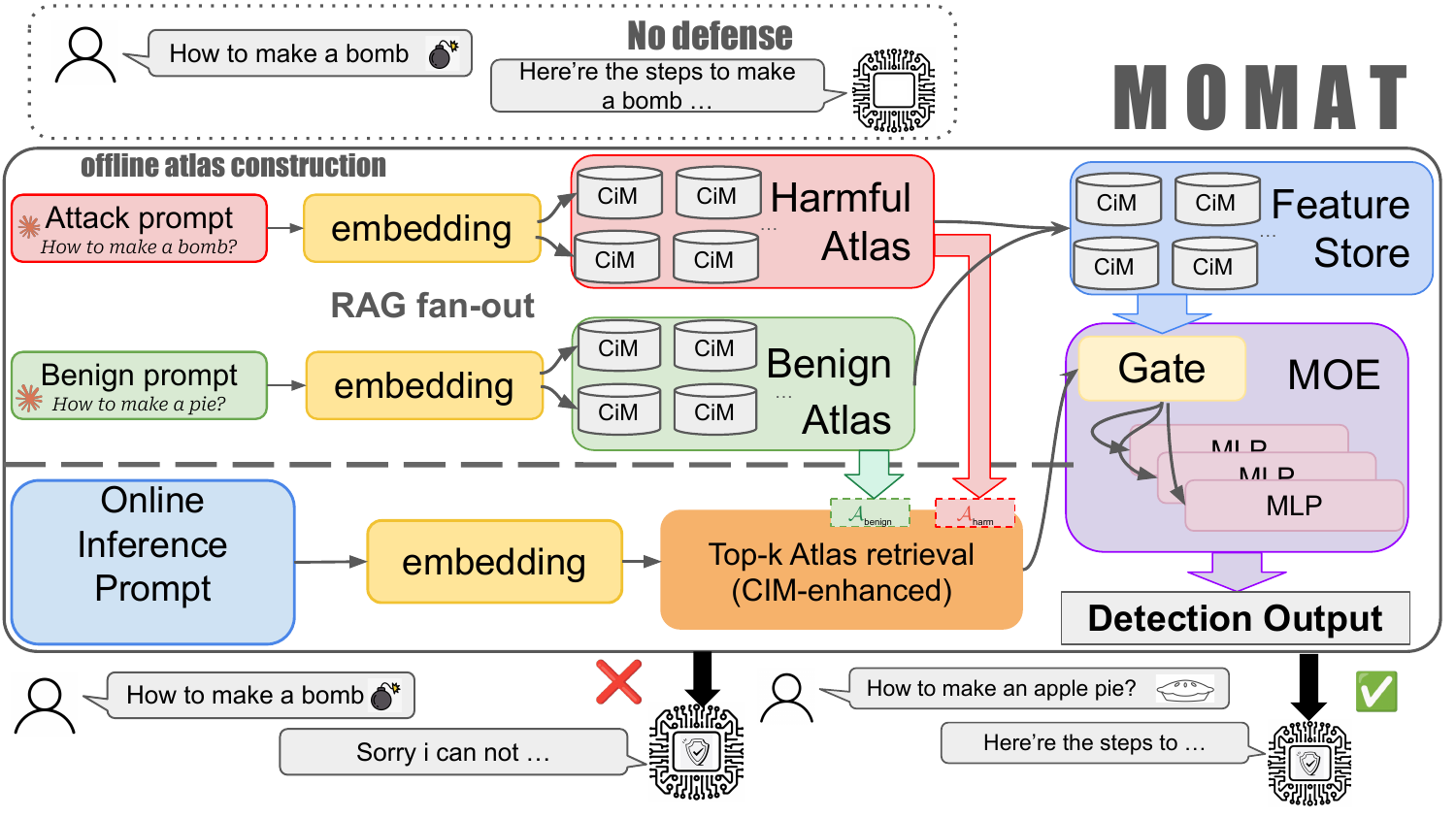}
\end{center}
   \caption{Overview of MOMAT. 
   The pipeline consists of two phases: an offline phase comprising (1) atlas construction from labeled examples and (2) training a lightweight MoE (1.03M parameters) with frozen embeddings, and an online inference phase where (3) MOMAT performs real-time multi-atlas CiM retrieval and MoE safety scoring on edge devices.
   }

\label{fig:demo}
\end{figure*} 
\sloppy
Quantized large language models (qLLMs) have become increasingly attractive for edge and embedded deployment because quantization significantly reduces model size, latency, and energy consumption \citep{b5, b29}. However, these advantages degrade alignment robustness \citep{b1}. Compressing model weights distorts the representation geometry and shrinks safety-relevant feature subspaces, while activation clipping during quantization further compounds these effects \citep{b2}. Together, these changes reduce the model's resistance to jailbreaks, prompt injection, and role-playing attacks \citep{b3}. Edge-deployed qLLMs are thus far more vulnerable than their full-precision cloud counterparts \citep{b33}, yet most run without dedicated safety mechanisms \citep{b31}. Efficient defense frameworks that operate within edge resource constraints are needed to close this gap.

Full-precision LLMs maintain robust safety boundaries through high-dimensional alignment features \citep{b4}, but these mechanisms degrade dramatically under aggressive quantization \citep{b8}. Quantization degrades safety alignment through several mechanisms: it introduces quantization errors from activation outliers \citep{b17}, disrupts the low-rank subspaces that encode safety-critical information \citep{b34}, and damages the ``safety layers" that help models distinguish malicious queries \citep{b35}. These effects compromise the model's ability to recognize and reject unsafe inputs, including role-playing scenarios, adversarially rephrased requests, and multi-turn steering attacks \citep{b36}. The vulnerability is heightened in edge deployments, where resource constraints limit the deployment of multi-model safety pipelines and server-side filtering that cloud environments rely on. Simple jailbreaks that full-precision models easily reject can succeed against quantized models running on edge devices. 
Beyond text-only alignment failures, our prior work shows that
linguistic pressure alone can induce hallucination in
vision-language models even without adversarial intent
\citep{b48, b49}, suggesting that compliance-driven fabrication is
a broader cross-modal phenomenon that similarly threatens the
reliability of quantized systems under structured or coercive
prompting.

Retrieval-augmented generation (RAG) can help mitigate jailbreaks by retrieving refusal templates or policy constraints \citep{b32}. But this approach struggles in large, heterogeneous safety databases. As the diversity of harmful intents grows, the embedding space becomes sparse. The curse of dimensionality means that semantically unrelated items end up at similar distances from the query. A safety-critical query may fail to retrieve the right exemplars and instead surface irrelevant or weakly related templates. This retrieval noise not only weakens downstream safety responses, but adversaries can also exploit it to bypass template matching entirely. Edge devices face an additional challenge: they lack the memory bandwidth for high-capacity vector search, which increases both latency and energy costs. A more structured approach is needed—one that uses domain-localized retrieval rather than a single monolithic database.

We introduce MOMAT (Mixture of Multiple Atlases), a hardware-accelerated safety framework that restructures the retrieval problem into multiple semantically coherent subspaces. Rather than using a single large safety database, MOMAT organizes harmful and benign samples and policy templates into specialized atlases. Each atlas captures a dense, thematically consistent region of the semantic space. This reduces cross-domain interference and improves nearest-neighbor reliability by operating within compact clusters. For each input prompt, the CiM engine retrieves top-$k$ similarity features from all atlases in parallel, enabling domain-localized retrieval with much less search noise. For edge deployment, MOMAT uses a CiM (Compute-in-Memory)-enhanced RAG engine that performs similarity search directly in memory arrays. This lowers both latency and energy consumption compared to DRAM-based vector retrieval. Together, MOMAT provides a modular safety architecture for defending qLLMs under edge constraints.

To summarize, this work makes three key contributions toward enabling reliable and energy-efficient jailbreak defense for quantized LLMs deployed on edge devices:

\begin{itemize}
\item We introduce MOMAT, a hardware-enhanced safety architecture that restructures safety retrieval into multiple semantically coherent atlases. This structured organization mitigates curse-of-dimensionality effects, reduces retrieval noise, and provides compact, reliable protective modules for edge-deployed qLLMs.
    
\item We develop a CiM–enhanced retrieval engine that performs similarity search directly within CiM, accelerating a 100-query batch from 15{,}052.44 ms to 3{,}207.21 ns (a $4.69\times10^{6}\!\times$ speedup) and reducing energy consumption from $8.1\times10^{7}\,\mu$J to $3.32\,\mu$J, yielding a $\sim2.5\times10^{5}\!\times$ energy reduction over Raspberry~Pi–class DRAM baselines.

\item We evaluate MOMAT on standard jailbreak benchmarks, demonstrating that it achieves state-of-the-art defense performance without introducing additional over-refusal of benign queries. To support future research, we will release the complete 223.2k-sample dataset.

\end{itemize}

\section{Background}
\label{sec:background}

\subsection{Quantization and Alignment Collapse}
\label{sec:background:alignment_collapse}

Quantization is a widely used technique for accelerating LLM inference on memory- and energy-constrained edge hardware \citep{b5}. By reducing numerical precision through weight compression and activation quantization, qLLMs achieve substantial reductions in model size, inference latency, and power consumption. However, these efficiency gains come at the cost of reduced representational fidelity \citep{b19}. Quantization disrupts high-dimensional feature geometry by introducing weight rounding errors and activation outlier clipping, which compress semantic distances and flatten safety-critical gradients in deeper transformer layers \citep{b17}. 
As a result, the low-rank subspaces that encode safety-relevant information are degraded, making it harder for the model to separate harmful from benign intent in embedding space \citep{b34}. The safety layers responsible for recognizing and rejecting adversarial inputs are disproportionately affected, as they rely on fine-grained representational distinctions that aggressive quantization erodes \citep{b35}. 

Quantized models are consequently far more susceptible to jailbreaks \citep{b18}, prompt injection, and multi-turn steering compared to their full-precision counterparts \citep{b20}. This vulnerability is compounded in edge deployments, where qLLMs typically run in isolation without auxiliary safety models, server-side monitoring, or high-bandwidth retrieval modules. Alignment collapse under quantization is therefore a critical and underaddressed vulnerability in real-world edge scenarios.

\subsection{Jailbreak and Prompt Injection Threat}
\label{sec:background:Threat}

Jailbreak and prompt-injection attacks coerce LLMs into generating harmful or 
policy-violating content by exploiting the model's reliance on natural-language cues rather than formal safety guarantees \citep{b7, b9}. These attacks span a broad taxonomy: role-playing and persona hijacking \citep{b14}, multi-turn context steering, translation-based obfuscation \citep{b12}, and adversarially paraphrased instructions \citep{b16}. Prompt-injection variants can override system messages, invert safety rules, or hijack in-context exemplars, effectively reprogramming the model's behavior at inference time without modifying its 
weights.

Quantized LLMs are especially vulnerable to these attacks. Weight and activation 
compression reduces robustness to linguistic variation, causing semantically adversarial inputs to be misclassified as benign in the distorted embedding space. Crucially, most jailbreak attacks require no privileged access and can be launched through standard user-facing interfaces, making them a practical and low-barrier threat for edge-deployed qLLMs that lack server-side filtering or safety oversight.

\subsection{Limitations of Safety RAG: Curse of Dimensionality and Semantic Sparsity}
\label{sec:background:Curse}

RAG is commonly used to improve LLM safety by retrieving refusal templates, safety 
exemplars, or policy snippets at inference time \citep{b9}. However, this approach degrades as the safety database grows large and semantically heterogeneous. As harmful-intent prompts diversify across categories, the embedding space grows increasingly sparse. The curse of dimensionality causes semantically unrelated vectors to converge to similar distances from a query point, undermining nearest-neighbor reliability \citep{b11}. Safety-critical queries consequently fail to retrieve the correct refusal exemplars and instead surface irrelevant, weakly related, or overly generic templates that are ineffective against nuanced jailbreak attempts.

The problem is further compounded when exemplars from semantically distant domains---such as self-harm, illicit instructions, harassment, and privacy leakage---coexist in a shared embedding index. This mixture introduces cross-domain interference that amplifies retrieval noise and degrades the precision of template-based safety mechanisms. Beyond retrieval quality, large monolithic RAG indices impose significant memory bandwidth demands: DRAM-based vector search requires repeated high-latency data transfers that are incompatible with the real-time safety enforcement requirements of edge hardware. Monolithic RAG designs are therefore poorly suited for qLLM safety on resource-constrained platforms, motivating a more structured, domain-localized retrieval approach.

\subsection{Edge Deployment Constraints and CiM-Based Accelerators}
\label{sec:background:cim}

Edge and embedded platforms impose stringent constraints on memory capacity, bandwidth, and energy consumption \citep{b21}, making it difficult to deploy safety mechanisms that depend on large retrieval indices or compute-heavy classifiers. DRAM-based vector search is particularly costly: each similarity query requires multiple high-latency memory accesses \citep{b23, b24}, and the energy spent moving data between memory and compute units dominates the overall retrieval cost \citep{b22}. Under the traditional von Neumann architecture, this memory wall prevents edge devices 
from sustaining the retrieval throughput that real-time safety enforcement demands \citep{b28}.

Compute-in-Memory (CiM) architectures offer a principled solution by performing Multiply-Accumulate (MAC) operations directly within memory arrays, eliminating the dominant source of data movement overhead \citep{b25, b27, b13}. CiM is particularly well-suited for retrieval workloads because nearest-neighbor search reduces to vector-matrix multiplication: query embeddings are broadcast as input vectors, the stored exemplar matrix resides in the crossbar array, and similarity scores are produced in a single analog MAC sweep without transferring data off-chip. This architectural alignment between retrieval computation and CiM hardware enables orders-of-magnitude reductions in both latency and energy per query compared to DRAM-based designs \citep{b28, b26}.

MOMAT exploits this property by storing atlas-specific exemplar matrices directly in dedicated CiM arrays, enabling fast, low-power similarity search over compact, domain-localized subspaces. This combination of structured retrieval and in-memory computation provides an effective hardware foundation for meeting the real-time latency and energy budgets required by edge-deployed safety systems.

\section{Method}
\label{sec:method}

\subsection{System Overview}
\label{sec:method:overview}

MOMAT is a CiM–assisted safety framework that restores alignment behaviors sacrificed in quantized LLMs deployed on edge hardware. MOMAT decouples prompt encoding from the protected qLLM: input prompts are encoded by a lightweight, encoder-only embedding model ($\mathcal{E}$; we use BAAI/bge-large-en-v1.5 \citep{b47}, 0.3B parameters), which is frozen throughout and serves solely to produce stable, full-precision embeddings for safety processing. This component can be replaced by any embedding model and is independent of the qLLM being defended. Rather than querying a domain-agnostic retrieval corpus, MOMAT organizes safety knowledge into multiple semantic atlases, enabling localized retrieval and reducing interference from unrelated domains. As shown in Fig.~\ref{fig:demo}, the system comprises four modules: 
a CiM retrieval module that computes top-$k$ similarity features from each atlas for every incoming query,
atlas-specific safety stores that maintain dense CiM arrays of harmful exemplars and benign counterparts organized into semantically coherent clusters, 
a lightweight MoE (Mixture of Experts) intent detector that fuses atlas-retrieved features with the prompt embedding to produce a safety decision, 
and a CiM-accelerated RAG engine that serves two roles: computing atlas-local top-$k$ similarity features for every query to support MoE safety scoring, and retrieving safe response templates directly within memory arrays when a harmful prompt is detected---both performed with low latency and energy suitable for edge hardware.

Given a prompt, MOMAT encodes it using the frozen embedding model $\mathcal{E}$ \citep{b47} and 
feeds the resulting 1024-dimensional embedding to the CiM engine, which retrieves top-$k$ exemplars from each of the $M$ atlases. Each atlas maintains dense clusters of harmful exemplars, policy snippets, and ICL (In-Context Learning)-generated variants organized within a dedicated CiM array. This structured atlas-local retrieval confines downstream processing to semantically relevant regions and suppresses cross-domain retrieval noise. A compact MoE detector then receives the prompt embedding concatenated with atlas-local top-$k$ precomputed features, and produces a binary safety decision. When a harmful intent is detected, MOMAT enforces a safety response; otherwise, the prompt is passed through to the underlying qLLM. 
Throughout inference, the CiM engine operates in two modes: it computes atlas-local top-$k$ similarity features $r_i(x)$ for every query to support MoE scoring, and retrieves the safe response template from the most activated atlas when harmful intent is confirmed, both entirely within CiM arrays and without DRAM-based data movement, enabling sub-millisecond, low-energy retrieval suitable for edge deployment.

This combination of domain-localized retrieval and CiM-based acceleration yields a modular defense pipeline that preserves the computational compactness of qLLMs while reinstating reliable safety enforcement. MOMAT amortizes the cost of safety reasoning across two phases: an offline atlas construction phase that handles all expensive operations including safety corpus processing, semantic clustering, and multi-stage ICL expansion, and a lightweight online inference phase that reduces to a single embedding computation followed by a compact MoE forward pass. At inference time, the CiM arrays produce atlas-local top-$k$ similarity signals in-situ, eliminating the need for DRAM-based vector search. This offline–online separation is key to meeting the latency and energy budgets of edge platforms, where neither large safety models nor high-bandwidth memory are available.

\subsection{Atlas Construction}
\label{sec:method:atlas}

MOMAT constructs multiple \emph{atlases} to replace the flat, domain-agnostic vector store used in conventional RAG systems. The enrichment yields dense neighborhoods well suited for stable similarity search on CiM systems.

Algorithm~\ref{alg:atlas-construction} builds atlases in three steps:
(1) use a small set of seed prompts to ICL-generate cross-domain harmful–benign pairs;
(2) recursively fan out each pair with paraphrasing, syntactic variation, semantic abstraction, and detailed expansions to obtain dense variant sets;
(3) assemble these variants into atlas-specific corpora, yielding domain-labeled, high-density neighborhoods.

These high-density neighborhoods reduce retrieval noise and align naturally with CiM-executed similarity search. Partitioning the corpus into multiple atlases also avoids high-dimensional interference that occurs when semantically unrelated domains share a single retrieval index, allowing atlas-local RAG to outperform monolithic retrieval in both accuracy and efficiency.

Atlas embeddings and statistics are precomputed offline, leaving only prompt encoding, CiM atlas retrieval, and MoE scoring for the online stage.

\begin{algorithm}[ht]
\caption{ICL-Based Atlas Construction for MOMAT}
\label{alg:atlas-construction}
\footnotesize
\begin{algorithmic}[1]

\Require Seed set $\mathcal{S}_0$; ICL generator $\mathcal{G}$; number of atlases $M$.
\Ensure Atlases $\{\mathcal{A}_1,\dots,\mathcal{A}_M\}$ with domain-labeled variant sets.

\Statex \textbf{Stage 1: Cross-domain fan-out}
\State Initialize $\mathcal{S}_1 \gets \emptyset$
\For{each seed $s \in \mathcal{S}_0$}
    \State $\mathcal{P}_s \gets \mathcal{G}\_\text{cross-domain}(s)$ 
    \Comment{Generate harmful–benign pairs across domains}
    \State $\mathcal{S}_1 \gets \mathcal{S}_1 \cup \mathcal{P}_s$
\EndFor

\Statex \textbf{Stage 2: Recursive variant expansion}
\State Initialize $\mathcal{S}_2 \gets \emptyset$
\For{each pair $p \in \mathcal{S}_1$}
\State $\mathcal{V}_p \gets \mathcal{G}_{\text{variants}}(p)$ \Comment{Variant expansion}

    \State $\mathcal{S}_2 \gets \mathcal{S}_2 \cup \mathcal{V}_p$
\EndFor

\Statex \textbf{Stage 3: Atlas assembly}
\State Initialize $\mathcal{A}_1,\dots,\mathcal{A}_M$
\For{each sample $x \in \mathcal{S}_2$}
    \State $i \gets \textsc{DomainLabel}(x)$   
    \Comment{Domain label is inherent from the fan-out path}
    \State $\mathcal{A}_i \gets \mathcal{A}_i \cup \{x\}$
\EndFor

\State \Return $\{\mathcal{A}_1,\dots,\mathcal{A}_M\}$

\end{algorithmic}
\end{algorithm}

\subsection{Atlas Retrieval and Lightweight Intent Detection}
\label{sec:method:routing-mlp}

After encoding a prompt, MOMAT feeds the prompt embedding $e_x$ to the CiM engine, which retrieves atlas-local top-$k$ similarity features $r_i(x)$ from each of the $M$ atlases against their precomputed exemplar embeddings. This confines safety processing to appropriate domains and is robust to paraphrasing, obfuscation, and role-play variations because routing depends on embedding-space geometry rather than surface phrasing. Within the routed set, MOMAT applies a small MoE module composed of several MLP experts. 
This module consumes both the prompt embedding and its atlas-local top-$k$ retrieval features, which are prepared offline during atlas construction. 

Routing and local scoring incur minimal cost: centroid comparisons are low-dimensional, and the experts in the MoE are implemented as lightweight MLPs.
At deployment, these components select which offline-prepared atlas-local features are forwarded to the main MoE scorer (Sec.~\ref{sec:method:moe}). This early localization stage keeps online computation small enough for edge hardware while preserving reliable safety behavior.

\subsection{CiM-Enhanced RAG Engine}
\label{sec:method:cim}

MOMAT stores each atlas as a compact vector matrix preloaded into CiM memory arrays, and the CiM engine serves two roles in the inference pipeline. First, during MoE safety scoring, the CiM engine retrieves atlas-local top-$k$ similarity 
features $r_i(x)$ from all $M$ atlases in parallel for every incoming query, providing the retrieval context that the MoE detector requires to produce a reliable safety decision. Second, when the MoE scorer determines that $p_{\mathrm{harm}} \geq \tau$, the CiM engine performs a targeted second-pass retrieval from the most activated atlas $\mathcal{A}_{i^\star}$ to surface a safe response template. In both cases, the CiM engine executes parallel MAC (Multiply-Accumulate) operations directly within the storage arrays, computing dot-product similarities without moving data off-chip. Unlike DRAM-based retrieval---which repeatedly shuttles vectors between memory and compute units, incurring significant bandwidth and energy overhead---CiM returns top-$k$ matches with minimal data movement. Because retrieval operates over preloaded exemplar matrices resident in the CiM arrays, the engine sustains low latency and energy consumption across both use cases.

\subsection{End-to-End Inference Pipeline}
\label{sec:method:inference}

Algorithm~\ref{alg:momat-inference} summarizes MOMAT's end-to-end inference pipeline. Given a prompt $x$, MOMAT first computes its embedding $e_x = \mathcal{E}(x)$ using the frozen encoder-only embedding model, 
then feeds $e_x$ to the CiM engine, which retrieves atlas-local top-$k$ similarity summaries $\{r_i(x)\}_{i=1}^{M}$ from all $M$ atlases in parallel,
confining all subsequent computation to these semantically coherent subspaces.

For each routed atlas $\mathcal{A}_i$, the system computes an atlas-local retrieval summary $r_i(x)$ via \textsc{LocalRetrievalSummary}, which aggregates top-$k$ similarity scores and descriptive statistics between $e_x$ and the atlas exemplars. 
Each of the $K$ independent MLP experts $f_j$ takes $\mathbf{z}(x)$ as input and produces a harmfulness logit $s_j(x)$. 
Simultaneously, a learned gating network produces unnormalized logits $\{\tilde{g}_j(x)\}_{j=1}^K$, which are normalized via softmax to yield mixture weights $\alpha_j(x)$. The final risk estimate is the gated aggregation $S(x) = \sum_{j=1}^{K} \alpha_j(x) s_j(x)$, passed through a sigmoid to produce the harmfulness probability $p_{\mathrm{harm}}$.

If $p_{\mathrm{harm}} < \tau$, the prompt is deemed benign and forwarded directly to the underlying qLLM for generation. Otherwise, MOMAT identifies the most activated atlas $i^\star = \arg\max_{i \in \{1,\dots,M\}}\, r_i(x)$ and invokes the CiM engine to retrieve the top-$k$ nearest exemplars from $\mathcal{A}_{i^\star}$ entirely in-situ. A safe response template $t_{\mathrm{safe}}$ is selected from the retrieved neighbors and used by \textsc{SafeCompose} to produce a policy-compliant response conditioned on both the original prompt and the qLLM.

Because MoE scoring operates over atlas-local top-$k$ features computed in parallel across all $M$ atlases, the online inference cost reduces to a single embedding computation, a CiM retrieval pass across all $M$ atlases to compute $r_i(x)$, a lightweight MoE forward pass, and---when $p_{\mathrm{harm}} \geq \tau$---a second targeted CiM retrieval pass from $\mathcal{A}_{i^\star}$ for safe template selection. All atlas exemplar embeddings are precomputed offline and remain resident in the CiM arrays, so neither CiM pass incurs DRAM data movement at inference time. This design ensures that MOMAT meets the latency and energy constraints of edge deployment while restoring the alignment behaviors degraded by quantization.

\begin{algorithm}[t]
\caption{MOMAT Inference Pipeline}
\footnotesize
\label{alg:momat-inference}
\begin{algorithmic}[1]

\Require Prompt $x$; frozen encoder-only embedding model 
$\mathcal{E}$; atlases $\{\mathcal{A}_i\}_{i=1}^{M}$; 
MLP-expert MoE $\{f_j\}_{j=1}^{K}$; gating function $w$; 
CiM engine; protected qLLM.
\Ensure Safe response $y$.

\State $e_x \gets \mathcal{E}(x)$
\Comment{Embed prompt using frozen encoder}

\Statex \textbf{CiM Retrieval: Atlas Feature Extraction}
\For{$i = 1, \dots, M$}
    \State $r_i(x) \gets \textsc{LocalRetrievalSummary}(e_x, \mathcal{A}_i)$
    \Comment{CiM Use 1: online top-$k$ feature extraction}
\EndFor
\State $\mathbf{z}(x) \gets \textsc{BuildFeatureVector}(e_x, \{r_i(x)\}_{i=1}^{M})$
\Comment{Construct full feature vector over all $M$ atlases}

\Statex \textbf{MoE Safety Scoring}
\For{$j = 1, \dots, K$}
    \State $s_j(x) \gets f_j(\mathbf{z}(x))$
    \Comment{Expert $j$ scores over full feature vector}
    \State $\tilde{g}_j(x) \gets w_j^\top e_x + b_j$
    \Comment{Gating logit for expert $j$}
\EndFor
\State $g_j(x) \gets \textsc{Softmax}(\{\tilde{g}_j(x)\}_{j=1}^{K})$
\Comment{Normalize gating weights}
\State $S(x) \gets \sum_{j=1}^{K} g_j(x)\, s_j(x)$
\Comment{Weighted aggregation of expert scores}
\State $p_{\mathrm{harm}} \gets \sigma(S(x))$
\Comment{Harmfulness probability}

\If{$p_{\mathrm{harm}} < \tau$}
    \State \Return $\textsc{qLLMGenerate}(x)$
    \Comment{Benign prompt: pass through to qLLM}
\EndIf

\Statex \textbf{CiM Targeted Retrieval and Safe Response}
\State $i^\star \gets \arg\max_{i \in \{1,\dots,M\}}\, r_i(x)$
\Comment{Most activated atlas}
\State $\mathcal{N} \gets \textsc{CiMRetrieveTopK}(e_x, \mathcal{A}_{i^\star})$
\Comment{CiM Use 2: targeted template retrieval}
\State $t_{\mathrm{safe}} \gets \textsc{SelectTemplate}(\mathcal{N})$
\State \Return $\textsc{SafeCompose}(x, t_{\mathrm{safe}}, \textsc{qLLM})$

\end{algorithmic}
\end{algorithm}

\subsection{MOMAT for Safety Detection}
\label{sec:method:objective}

While the previous subsections describe MOMAT at the architectural level, we now formalize its multi-atlas comparison mechanism and training objective. Unlike traditional contrastive learning approaches that fine-tune embeddings, MOMAT treats the query embedding as \emph{fixed} (produced by the frozen encoder-only model $\mathcal{E}$, independent of the qLLM being defended) and instead learns to interpret the structural patterns revealed by retrieval from multiple reference atlases.

\subsubsection{Multi-Atlas Architecture.}
\label{sec:method:multi-atlas}

Let $x$ denote an input prompt and $e_x \in \mathbb{R}^d$ its embedding computed by the frozen encoder-only model $\mathcal{E}$. We maintain $M$ pre-constructed reference atlases, where each atlas $\mathcal{A}_i$ ($i = 1, \dots, M$) represents a distinct semantic domain relevant to the safety detection task. The design is flexible:
\begin{itemize}
    \item \textbf{Binary setup}: $M=2$ with harmful vs.\ benign atlases
    \item \textbf{Multi-domain setup}: $M > 2$ with fine-grained categories (e.g., violence, hate speech, self-harm, benign)
    \item \textbf{Hierarchical}: Combining coarse and fine-grained atlases
\end{itemize}

For each query $x$, we retrieve the top-$k$ most similar examples from each atlas $\mathcal{A}_i$ using CiM-enhanced similarity search (Sec.~\ref{sec:method:cim}). Let $\mathbf{s}_i(x) \in \mathbb{R}^k$ denote the vector of top-$k$ similarity scores from atlas $i$.

Rather than using these scores directly, we construct a \emph{rich feature representation} $\mathbf{z}(x)$ that captures:
\begin{enumerate}
    \item The original query embedding: $e_x \in \mathbb{R}^d$
    \item Raw similarity distributions: $\mathbf{s}_1(x), \dots, \mathbf{s}_M(x) \in \mathbb{R}^k$
    \item Statistical summaries per atlas ($i = 1, \dots, M$):
    \begin{equation}
        \mu_i = \text{mean}(\mathbf{s}_i),\quad 
        \max_i = \max(\mathbf{s}_i),\quad 
        \min_i = \min(\mathbf{s}_i)
    \end{equation}
    \item Pairwise comparative features for atlas pairs $(i,j)$:
    \begin{equation}
    \begin{aligned}
        \Delta_{ij}^{\mu} &= \mu_i - \mu_j, \quad
        \Delta_{ij}^{\max} = \max_i - \max_j, \quad
        \rho_{ij} = \frac{\mu_i}{\mu_j + \epsilon}
    \end{aligned}
    \end{equation}
\end{enumerate}

The complete feature vector is:
\begin{equation}
\begin{split}
    \mathbf{z}(x) = \Big[&\, e_x;\; \{\mathbf{s}_i\}_{i=1}^M;\; \{\mu_i, \max_i, \min_i\}_{i=1}^M; \\
    &\, \{\Delta_{ij}^{\mu}, \Delta_{ij}^{\max}, \rho_{ij}\}_{i<j}\Big] \in \mathbb{R}^D,
\end{split}
\end{equation}
where $D = d + M \cdot k + 3M + 3\binom{M}{2}$. This construction generalizes to arbitrary $M$ while maintaining interpretability through explicit comparative statistics.

\subsubsection{Mixture-of-Experts for Pattern-Specific Routing.}
\label{sec:method:moe}

Different queries exhibit distinct \emph{neighborhood patterns} across the multi-atlas space. To handle this diversity, we employ a Mixture-of-Experts (MoE) architecture where each expert $f_j(\cdot)$ specializes in a particular neighborhood signature. Specifically, we define $K$ expert networks (typically $K = 3$ 
to 5), each implemented as a multi-layer perceptron:
\begin{equation}
    s_j(x) = f_j(\mathbf{z}(x)), \quad j = 1, \dots, K,
\end{equation}
where $s_j(x) \in \mathbb{R}$ is expert $j$'s logit score for the target safety prediction.

A gating network, conditioned only on the query embedding $e_x$, produces expert routing weights:
\begin{equation}
    \alpha_j(x) = \frac{\exp(w_j^\top e_x + b_j)}
    {\sum_{j'=1}^K \exp(w_{j'}^\top e_x + b_{j'})},
\end{equation}
where $\{w_j, b_j\}_{j=1}^K$ are learnable gating parameters.

The final logit is computed as:
\begin{equation}
    S(x) = \sum_{j=1}^K \alpha_j(x) \, s_j(x),
\end{equation}
with the predicted probability given by 
$p(y=1 \mid x) = \sigma(S(x))$.

\paragraph{Training Objective.}
Given training data $\mathcal{D} = \{(x_n, y_n)\}_{n=1}^N$, the primary objective is binary cross-entropy:
\begin{equation}
\begin{split}
    \mathcal{L}_{\text{cls}} = - \frac{1}{N} \sum_{n=1}^{N} \big[
        y_n \log p(y_n\!=\!1 \mid x_n) \\
        + (1-y_n) \log\big(1 - p(y_n\!=\!1 \mid x_n)\big)
    \big].
\end{split}
\end{equation}

To prevent expert collapse, we add a gating balance regularizer:
\begin{equation}
    \mathcal{L}_{\text{balance}} = \frac{1}{K} \sum_{j=1}^K 
    \left( \bar{\alpha}_j - \frac{1}{K} \right)^2,
\end{equation}
where $\bar{\alpha}_j = \frac{1}{N}\sum_{n=1}^N \alpha_j(x_n)$.

The overall training objective is:
\begin{equation}
    \mathcal{L} = \mathcal{L}_{\text{cls}} + 
    \lambda_{\text{balance}} \mathcal{L}_{\text{balance}}.
\end{equation}

This formulation naturally scales from binary ($M=2$) to multi-domain ($M>2$) scenarios, with the MoE architecture adapting to the increased feature complexity as $M$ grows.

\section{Experiment}
\label{sec:experiment}

This section evaluates MOMAT under realistic edge deployment conditions and quantized LLM settings. Our goal is to characterize three aspects of the system:
(1) Safety robustness, i.e., whether MOMAT recovers the alignment capabilities that quantization erodes;
(2) Deployment efficiency, i.e., whether CiM-based in-memory retrieval achieves the latency and energy targets required for edge inference; and
(3) Utility preservation, whether MOMAT maintains benign performance without over-blocking.
To cleanly separate algorithmic improvements from hardware acceleration, we conduct experiments across two hardware environments and multiple qLLM configurations, using multilingual jailbreak and prompt-injection benchmarks that correspond directly to MOMAT’s atlas taxonomy.

\subsection{Evaluation Settings}

We evaluate MOMAT across hardware, model, and task conditions that reflect realistic edge deployment. To separate algorithmic gains from hardware acceleration, we use two environments. On the Raspberry Pi 5, we run the full MOMAT pipeline (CiM atlas retrieval and MoE intent detection) while keeping retrieval DRAM-bound. 
This configuration isolates MOMAT's structural advantages, specifically atlas-localized retrieval versus monolithic RAG, under the memory bandwidth and latency constraints typical of CPU-only embedded platforms.  On the CiM evaluation board, we preserve the identical MOMAT control flow but execute retrieval through a 4-bit/8-bit in-memory MAC array, isolating the benefit of CiM-accelerated similarity search without introducing confounding algorithmic changes. Together, these two platforms probe MOMAT's two primary design axes: semantic retrieval reliability (Raspberry Pi) and energy-efficient in-memory retrieval (CiM).

We experiment with two representative qLLMs—Llama-2-7B and Mistral-7B—covering common quantization regimes used for edge inference. These models display different degrees of representation distortion and alignment collapse under quantization, allowing us to assess how effectively MOMAT compensates for safety degradation across varying compression levels.

To assess safety performance, we conduct red-team evaluations using domain-specific harmful instruction sets spanning categories such as weapons, self-harm, and privacy violations. These categories are aligned with the atlas taxonomy introduced in Sec.~\ref{sec:method:atlas}, enabling fine-grained analysis of MOMAT's domain-localized retrieval and its ability to surface contextually appropriate safety exemplars for each threat category.

We report four metrics: ASR (Attack Success Rate), FRR (False Refusal Rate), p95 end-to-end latency, and energy per query. ASR measures the fraction of adversarial prompts that successfully elicit a harmful response from the defended model, a lower ASR indicates stronger jailbreak resistance. FRR measures the fraction of benign queries incorrectly refused by the safety system, capturing the over-refusal problem that plagues many conservative defenses; a lower FRR indicates that the system preserves normal qLLM utility for legitimate users. Latency and energy per query reflect MOMAT's hardware efficiency objectives. Together, these four metrics provide a holistic view of MOMAT's core design goals: restoring safety degraded by quantization, preserving qLLM usability on benign inputs, and satisfying the real-time, low-power constraints of edge deployment.

\begin{table}[t]
\centering
\fontsize{8pt}{10pt}\selectfont
\caption{Impact of Quantization on Model Size and ASR}
\label{tab:size_quantization1}
\resizebox{\columnwidth}{!}{%
\begin{tabular}{llccccc}
\toprule
 & & \multicolumn{2}{c}{\textbf{Size (GB)}} & \multicolumn{3}{c}{\textbf{ASR (\%)}} \\
\cmidrule(lr){3-4} \cmidrule(lr){5-7}
\textbf{Model} & \textbf{Dataset} & \textbf{FP16} & \textbf{W4A8} & \textbf{FP16} & \textbf{W4A8} & \textbf{$\Delta$} \\
\midrule
Llama2-13B & AdvBench & 26.03 & 7.00 & 9.0 & 8.3 & -0.7 \\
Llama2-7B & AdvBench & 13.38 & 3.76 & 28.1 & 32.5 & +4.4 \\
\midrule
Llama2-13B & Malicious & 26.03 & 7.00 & 21.0 & 19.0 & -2.0 \\
Llama2-7B & Malicious & 13.38 & 3.76 & 25.0 & 28.0 & +3.0 \\
\bottomrule
\end{tabular}%
}
\end{table}


\subsection{Results and Analysis}

\subsubsection{Safety Collapse of Small Quantized Models}
We first characterize how quantization affects safety alignment across model scales. Table \ref{tab:size_quantization1} reports model size and ASR for Llama2-7B and Llama2-13B before and after W4A8 \citep{b42} quantization, evaluated on both AdvBench \citep{b43} and Malicious Instruct \citep{b44} datasets. The results reveal a clear scale-dependent divergence. The 13B model exhibits negligible ASR change under quantization ($\Delta=-0.7\%$ on AdvBench, $\Delta=-2.0\%$ on Malicious Instruct), suggesting that larger models retain sufficient representational capacity to absorb quantization-induced perturbations without catastrophic alignment loss. In contrast, the 7B model suffers consistent ASR increases across both datasets ($\Delta=+4.4\%$ on AdvBench, $\Delta=+3.0\%$ on Malicious Instruct), indicating that smaller models lack the redundancy in their safety-relevant feature subspaces to withstand aggressive weight compression. 
This scale-dependent vulnerability carries significant implications for edge deployment. Despite its robustness, the 13B model is impractical for most edge hardware due to its 26 GB footprint in FP16, which far exceeds the memory capacity of embedded platforms. The 7B model, which compresses to 3.76 GB under W4A8 and is therefore the natural choice for edge deployment, is precisely the scale most susceptible to alignment collapse. This creates a fundamental tension: the compression techniques required to make LLMs deployable on edge hardware are the same techniques that undermine their safety alignment.

These findings directly motivate MOMAT's design. When a 7B qLLM undergoes W4A8 quantization, the resulting distortion of its embedding geometry degrades the separation between harmful and benign intents in feature space. A monolithic RAG database cannot reliably surface the correct safety exemplars under these conditions, as retrieval quality depends on the very geometric structure that quantization erodes. MOMAT addresses this by organizing safety knowledge into domain-localized atlases, each spanning a compact and semantically coherent subspace. By confining retrieval to these structured subspaces rather than querying across a heterogeneous corpus, MOMAT recovers reliable safety guidance even when the underlying qLLM has lost the global embedding structure needed for robust alignment.

\begin{figure}[pos=t]
    \centering
    \includegraphics[width=\columnwidth]{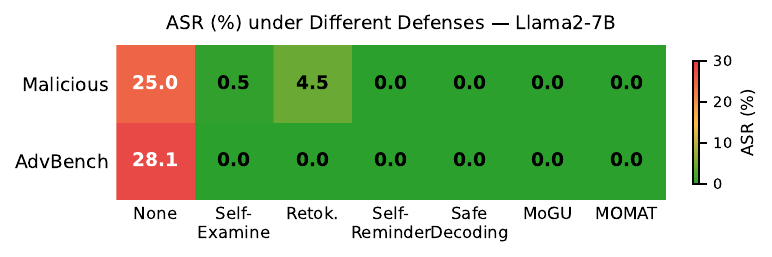}
    \caption{Attack Success Rate (ASR) of different defense methods on the Llama2-7B model (W4A8). Green indicates near-zero ASR; red indicates high vulnerability without defense.}
    \label{tab:asr_results}
\end{figure}

\begin{table}[t]
\fontsize{9pt}{10pt}\selectfont
\centering
\caption{ASR and FRR Results on W4A8 Quantized Models}
\label{tab:momat_comparison}
\resizebox{\columnwidth}{!}{%
\begin{tabular}{llcccc}
\toprule
 & & \multicolumn{2}{c}{\textbf{AdvBench}} & \multicolumn{2}{c}{\textbf{Malicious}} \\
\cmidrule(lr){3-4} \cmidrule(lr){5-6}
\textbf{Model} & \textbf{Defense} & \textbf{ASR} & \textbf{FRR} & \textbf{ASR} & \textbf{FRR} \\
\midrule
Llama2-7B & - & 32.5 & 13.3 & 28.0 & 14.0 \\
Llama2-7B & MOMAT & 0.00 & 13.3 (+0.0) & 0.00 &14.0 (+0.0) \\
\midrule
Mistral-7B & - & 68.2 & 12.7 & 67.5 & 17.0 \\
Mistral-7B & MOMAT & 0.00 & 12.7 (+0.0) & 0.0 & 17.0 (+0.0) \\
\bottomrule
\end{tabular}%
}
\end{table}

\subsubsection{MOMAT Safety performance}
Figure~\ref{tab:asr_results} compares ASR across defense methods on the W4A8-quantized Llama2-7B model. Without any defense, the model is highly vulnerable, with ASR reaching 28.1\% on AdvBench and 25.0\% on Malicious Instruct. Most defenses reduce ASR substantially, but with notable variation: Retokenization (Retok.) still permits 4.5\% ASR on Malicious Instruct, and Self-Examine leaves 0.5\% on the same dataset. MOMAT, along with SelfReminder, SafeDecoding, and MoGU, achieves 0.0\% ASR across both datasets. Critically, unlike methods such as SafeDecoding that rely on multi-step decoding or repeated model queries, MOMAT achieves this without token-level intervention at generation time, making it substantially more practical for latency- and power-constrained edge deployments.


Table~\ref{tab:momat_comparison} further examines MOMAT's behavior under W4A8 quantization across both Llama2-7B and Mistral-7B. Mistral-7B exhibits significantly higher baseline vulnerability, with ASR reaching 68.2\% on AdvBench and 67.5\% on Malicious Instruct under no defense, substantially worse than Llama2-7B (32.5\% and 28.0\%, respectively), consistent with Mistral's weaker alignment pretraining relative to Llama2's RLHF-tuned safety layers. MOMAT reduces ASR to 0.0\% on both models and both datasets, demonstrating robust defense generalization across model families and quantization-induced alignment collapse of varying severity.

Equally important, MOMAT introduces no additional over-refusal on benign queries. FRR remains unchanged after applying MOMAT on both models, confirming that the domain-localized atlas design avoids the conservative over-blocking that commonly afflicts safety systems relying on coarse-grained prompt filtering. The benign queries used for FRR evaluation were synthesized by GPT-4o-mini to share the same topic as their corresponding adversarial queries while requesting only harmless information—a deliberately conservative test of MOMAT's ability to distinguish intent within semantically proximate query pairs.

\begin{table}[t]
\fontsize{8pt}{10pt}\selectfont
\centering
\caption{DRAM window sweep on Raspberry Pi~5 for 100-query batches (162{,}084 vectors, 1024-D, top-$k=10$). Latency is averaged over 10 runs.}
\label{tab:rpi-dram-sweep}
\resizebox{\columnwidth}{!}{%
\begin{tabular}{rccc}
\toprule
DRAM window & DRAM size & Chunks & Avg time \\
(vectors) & (MB) & (\#) & (ms / 100 queries)\\
\midrule
   500  &  1.95 & 325 & 15{,}052.44 \\
 1{,}000 &  3.91 & 163 & 15{,}763.70 \\
 2{,}000 &  7.81 &  82 & 15{,}788.72 \\
 4{,}000 & 15.62 &  41 & 15{,}827.99 \\
10{,}000 & 39.06 &  17 & 15{,}911.68 \\
30{,}000 &117.19 &   6 & 16{,}034.05 \\
\midrule
\textbf{Optimal} & \textbf{1.95} & \textbf{325} & \textbf{15{,}052.44} \\
\bottomrule
\end{tabular}%
}
\end{table}

\subsubsection{Retrieval Latency and Energy Comparison: RPi DRAM Retrieval vs. MOMAT-CiM}


\paragraph{Raspberry Pi 5 DRAM Baseline.}

Table~\ref{tab:rpi-dram-sweep} reports CPU/DRAM retrieval latency on a Raspberry Pi~5 over a 162{,}084-vector safety database (1024-D, top-$k\!=\!10$) using 100-query batches. The optimal DRAM window is 500 vectors ($\approx\!1.95$~MB), yielding 15{,}052.44~ms per batch (150.52~ms per query, 6.6~QPS). Counterintuitively, increasing the DRAM window monotonically degrades latency---reaching 16{,}034.05~ms at 30{,}000 vectors---due to growing cache pressure and memory traffic overhead that outweighs the reduction in chunking iterations. For a larger 500-query batch, DRAM-based retrieval requires 84{,}681.67~ms (162.85~ms per query, 6.14~QPS), confirming that throughput scales poorly with batch size under DRAM-bound conditions.

\paragraph{SRAM-based CiM}


\begin{table}[t]
\fontsize{9pt}{10pt}\selectfont
\centering
\caption{Latency and energy for Raspberry Pi~5 DRAM retrieval vs.\ SRAM-based CiM retrieval (1024-D, top-$k=10$). DRAM energy is not measured; CiM energy is reported from the circuit-level simulator.}
\label{tab:dram-vs-cim}
\resizebox{\columnwidth}{!}{%
\begin{tabular}{lccc}
\toprule
Setting & Latency & Throughput & Energy / query \\
\midrule
RPi 5 (100Q) 
  & 15{,}052.44~ms 
  & 6.6~QPS 
  & $\ge 8.1\times10^{7}\,\mu\text{J}$ \\
CiM (100Q) 
  & 3{,}207.21~ns 
  & $3.1\times 10^7$~QPS
  & 3.32~$\mu$J \\
CiM (500Q) 
  & 3{,}213.15~ns 
  & $1.6\times 10^8$~QPS 
  & 3.32~$\mu$J \\
\bottomrule
\end{tabular}%
}
\end{table}


Table~\ref{tab:dram-vs-cim} reports simulation results for the SRAM-based CiM engine configured for 1024-D vectors and top-$k\!=\!10$ search. For a 100-query batch, the CiM array completes retrieval in 3{,}207.21~ns with a total energy of 3{,}321{,}779.62~pJ ($3.32\,\mu$J). Scaling to a 500-query batch---a $5\times$ increase in workload---adds only 5.94~ns in latency and 1.49~pJ in energy, yielding throughput of $1.6\times10^{8}$~QPS. This near-constant scaling behavior reflects a fundamental property of CiM architecture: since database vectors remain resident in the array once programmed, each query triggers only a single analog MAC sweep whose completion time is determined by vector dimensionality and array configuration rather than batch size or corpus cardinality. This contrasts sharply with DRAM-based retrieval, where both latency and energy scale with the number of memory transfers. The dominance of peripheral access over compute time in CiM arrays has been well established in prior hardware studies \cite{b46}.


\paragraph{Energy Estimation.}

The energy gap between the two retrieval paradigms is substantial. The Raspberry Pi~5 consumes approximately $8.1\times10^{7}\,\mu$J for a 100-query batch under conservative board-level estimation, while the CiM array consumes $3.32\,\mu$J for the same workload---a reduction of approximately $2.5\times10^{5}\times$. We note that these figures are not directly comparable in absolute terms, as the Raspberry Pi estimate reflects full board-level power draw while the CiM figure is derived from circuit-level simulation. Nevertheless, the magnitude of the gap is unambiguous and robust to reasonable estimation uncertainty, demonstrating that CiM-based retrieval provides a qualitatively different efficiency operating point that is inaccessible to DRAM-bound embedded platforms.

\section{Conclusion}
\label{sec:conclusion}

Quantized LLMs enable efficient edge deployment but suffer significant alignment degradation, making them highly susceptible to jailbreak and prompt-injection attacks. Our experiments confirm that this vulnerability is scale-dependent: while 13B models absorb W4A8 quantization with negligible ASR change, 7B models---the only scale practically deployable on edge hardware---exhibit consistent alignment collapse across datasets. This creates a fundamental tension between the compression required for edge deployment and the representational capacity required for reliable safety alignment. Flat, domain-agnostic retrieval databases exacerbate this problem. Under the curse of dimensionality, sparse high-dimensional embedding spaces prevent monolithic RAG systems from reliably surfacing relevant safety exemplars, particularly when the underlying qLLM's embedding geometry has already been distorted by quantization.

MOMAT addresses both failure modes by restructuring safety retrieval into semantically coherent, domain-localized atlases and executing similarity search on a CiM-accelerated engine. By confining retrieval to compact subspaces, MOMAT preserves stable local neighborhoods even under quantization-induced embedding distortion. By offloading similarity search to CiM arrays, MOMAT achieves a $4.69\times10^{6}\times$ latency reduction and a $\sim2.5\times10^{5}\times$ energy reduction over DRAM-based baselines. Critically, MOMAT reduces ASR to $0.0\%$ across both Llama2-7B and Mistral-7B under W4A8 quantization without introducing any additional over-refusal ($\Delta\,\text{FRR} = +0.0\%$), demonstrating that strong safety enforcement and benign utility preservation are not mutually exclusive under edge constraints.

These results suggest that reliable safety for edge-deployed qLLMs requires co-designing the retrieval structure and the underlying hardware substrate---a principle that generalizes beyond MOMAT to the broader challenge of deploying trustworthy AI on resource-constrained devices. Future work will investigate adaptive atlas construction to handle evolving threat taxonomies, and extend MOMAT's evaluation to a wider range of quantization schemes and on-device deployment environments.

\clearpage
\bibliographystyle{plainnat}
\bibliography{reference}

\end{document}